# DESCRIBING RATES OF INTERACTION BETWEEN MULTIPLE AUTONOMOUS ENTITIES: AN EXAMPLE USING COMBAT MODELLING

M K Lauren (September 2001)

Defence Technology Agency

New Zealand Defence Force

Private Bag 32901

Devonport

Auckland

New Zealand

e-mail: m.lauren@dta.mil.nz

fax: +64-9-445 5902

phone: +64-9-445 5445

## ABSTRACT

This paper endeavors to show how a fractal-based approach can offer an alternative to differential equations for describing rates of interaction in complex systems. The specific example given is for combat analysis. This concerns a scenario of some similarity to the situation in Mogadishu in 1993, for which military analysts struggled to provide useful analytical tools. It is seen that the correlated dispositions of the entities in the model lend themselves to being described by fractal dimensions. This appears to lead to power laws for the rate of attrition. An equation is suggested to describe this phenomenon. This is particularly interesting because it is an example of a power-law dependent on one of the parameters of the model rather than a temporal or spatial scale.

# INTRODUCTION

## Overview

This paper endeavors to show how methods such as cellular automaton models and fractal equations can replace differential equations as methods for addressing real world problems. The origin of this work is in the area of military analysis, specifically, estimating combat losses.

The standard equation for estimating casualty rates is the Lanchester equation (Lanchester, 1914):

$$\begin{aligned} \frac{dR}{dt} &= -k_B B(t), \quad R(0) = R_0 \\ \frac{dB}{dt} &= -k_R R(t), \quad B(0) = B_0 \end{aligned} \qquad (1)$$

where $R$ and $B$ represent the numerical strength at time $t$ of opposing Red and Blue forces, and $k_B$ and $k_R$ the killing rate of a Red/Blue individual.

However, the Lanchester equations make a number of gross assumptions, in particular, that all the battlefield entities are homogenous and evenly distributed. Such an equation is virtually useless in many modern scenarios, where the battlefield is often much more dispersed and likely to exhibit spatial and temporal correlations of deployments.

A generic scenario is used here to illustrate this concept. A small but powerful Blue force must maneuver its way through a much larger but dispersed Red force. Figure 1 shows the appearance of the model. The entities in the model are divided into two groups with distinct personalities:

**Blue troops:**

- Try to move from the top right-hand side of the screen down to the bottom left-hand side of the screen.
- Stay with friends, but maintain a certain minimum spacing if possible.
- Avoid Red.
- Have firepower equal or superior to Red ($k_B = 0.2$).

**Red hostiles:**

- Move toward friends if in a group of less than 3.
- Move toward Blue when they see them.
- Red firepower is systematically varied from $k_R = 0.01$ to 0.2.
- Require a local (i.e. within detection range) numerical advantage of more than 3 to attack Blue.

Such a scenario may represent a range of real operations taken from events in recent years, in particular, the attempts of US troops to force their way through the streets of Mogadishu in 1993.

Here, we will consider three variations:

Case I: Baseline.

Case II: Red have the ability to communicate beyond visual range.

Case III: A stochastic Lanchester model. For this case, it was simply assumed that all Red and Blue could shoot at each other.

**The MANA model**

The model used to explore this scenario is the MANA model (Lauren, Stephen and Anderson, 2001), produced by New Zealand's Defence Technology Agency. MANA is a cellular automaton model, but with the key difference from traditional cellular automaton models that it allows global interactions.

Each automaton is governed by a set of weightings that determine its propensity to move toward/away from friends/enemies and goals or waypoints. These are modified by requirements for the automata to keep a minimum distance from these goals, or to only attack enemies when a certain numerical advantage exists, or to only advance in the company of a certain number of friends.

The model also records maps of the locations of enemies for each side. The entities can be allowed to react to such information, effectively representing communications between entities.

Also, each entities has a set of parameters which describe its basic capabilities, e.g. movement rate, firing range, and detection range.

For the scenario considered here, the parameters described above are the only ones used.

## ANALYSIS

### A fractal battlefield

Cellular automata models are known to produce power-law, or “fractal”, distributions, as highlighted by the work of Bak et al (1989), while the current crop of conventional combat models do not. This is particularly interesting given that there is evidence that power laws exist in real combat data (Moffat and Passman, 2001; Roberts and Turcotte, 1998; Richardson, 1941, 1960). We expect that the patterns into which the entities in the MANA model evolve are likely to be able to be characterized by a fractal power law. Such patterns can be characterized by a fractal dimension, $D$, defined as:

$$D = \lim_{d \to 0} \frac{\log N}{\log\left(\frac{1}{d}\right)} \qquad (2)$$

where $N$ is the number of boxes of side length $d$ that are required to completely cover the distribution (Mandelbrot, 1983). If for some range of $d$ the value for $D$ is non-integer, then the patterns may be said to behave as a fractal for that range.

Equation 2 is easily adapted to the MANA model, since we need simply count how many cells contain entities, then aggregate cells and count again, repeating this process several times. However, there are also difficulties in trying to obtain such a measure. For one, the distribution of the entities is constantly changing. We might try recording the locations of casualties and obtain a fractal dimension of this instead. But for a single run, the number of casualties provides a relatively sparse pattern to try to analyze. An alternative is to analyze the collected locations of multiple runs. Given a large enough set of runs, one would expect that eventually every cell would have at least one kill occur it in. Therefore, to use this method, we must use a fixed number of runs to generate patterns for each of the different cases.

Figure 2 shows all the locations of kills for 600 runs of case II, where we have set Red's $k$ to 0.1. Analyzing how the number of cells containing kill locations changes as the size of the cells increases produces a straight-line fit on a log-log plot, with slope of 1.40.

Given this characterization of the distribution of the forces during engagements, we can now postulate as to the attrition rate for Blue. If $B$ is the number of remaining Blue, and $R$ is the remaining number of Red, then the attrition rate should only depend on:

$$\frac{\Delta B}{\Delta t} = f(R, t, k, D) \qquad (3)$$

where $t$ is time.

As discussed in the introduction, the Lanchester equation does not take account of spatial and temporal correlations. The spatial correlations are described by $D$, but it is also reasonable that the degree of temporal correlations in casualties must be a function of *D*, since the clustering of entities directly affects the timing of engagements.

It is therefore postulated that not only does the spatial distribution display a power law dependence in *d*, but the temporal structure function also obeys a power-law:

$$\left\langle \left| B(t_0 + \Delta t) - B(t_0) \right|^2 \right\rangle \propto \Delta t^{F(D)} \qquad (4)$$

where we expect *F* (*D*) is a non-integer. In order for this equation to be dimensionally correct, the right-hand side needs to be multiplied by another power of a unit with dimensions of time.

From equation 3, only *k* has dimensions of time, suggesting an equation of the form:

$$\left\langle \left| B(t_0 + \Delta t) - B(t_0) \right| \right\rangle \propto R\, k^{F(D)}\, \Delta t^{F(D)} \quad (5)$$

Note that for a given run, the Red automata may or may not evolve into a pattern with a fractal dimension similar to the majority of the other runs. Suppose we are interested in how quickly the Blue casualty level reaches some pre-decided level. Depending on the value of *k*, typically only a certain ensemble of runs will reach this level. For this ensemble, the distribution into which the Red automata evolve must be sufficiently dense to allow this level of casualties to be caused. Hence, by selecting just this ensemble, it seems reasonable that we are also selecting runs with similar fractal dimensions for the distribution of forces. Thus, we add the condition to equation 5 that the angled brackets represents an average over the ensemble of runs which actually reached this pre-determined level of casualties.

**Dependence of attrition on *k***

The right-hand side of equation 5 suggests that the rate of Blue attrition should depend on *k* to a non-integer power-law. Figure 3 shows the attrition rate for Blue as a function of *k* for cases I, II and III. Each point on the plot was calculated from 600 runs. Note that the attrition rate was calculated by finding the mean time to reach four casualties, for just those cases that reached this level. Although not as convincing a fit as seen in some of our earlier studies, nevertheless a power-law seems to be a reasonable approximation. Note that since reducing Red firepower requires Red to cluster to a greater extent to do the same damage to Blue, one cannot expect the fractal dimension for the Red distribution to necessarily remain the same for all values of $k$. If $D$ is a function of $k$, then it would be reasonable to expect departure from a perfect power law.

It is notable that case II has a much steeper slope than case I, indicating that communications play an important role in Red concentrating its firepower, hence $k$ has greater influence on casualties.

Figure 4 plots the number of runs for which four casualties occur (out of 600). Here, there appear to be at least two distinct regions: i) the left-hand side of the graph, displaying a steep, approximately power-law, slope, and; ii) the right-hand side, which is “flat” and independent of $k$. Obviously the right-hand side represents situations where nearly all runs produce casualty levels of four or greater. As one moves to the left on the graph, the plotted points gently slide away from this level, indicating a slowly growing number of runs for which casualties are light.

At the left-hand side of the figure, the number of runs reaching four casualties falls dramatically, in this case, approximated by a power law, i.e.:

$$N(C \geq 4) \propto k^{1.88} \quad (6)$$

However, this slope depends very much on the points chosen and might vary anywhere from an exponent of 1 to 2. Indeed, it may be tempting to try to identify three regions, a left, right and central to which straight lines could be fitted. Nevertheless, equation 6 may make a useful first-order approximation.

These two cases display behavior significantly different from the Lanchester case (case III) in two important ways: i) for the ensemble of runs that do reach the four casualty level, the rate of attrition is less strongly dependent on *k* than for the Lanchester model of attrition; ii) only a certain portion of the runs reach this level.

The second point highlights the existence of fat-tailed distributions of outcomes for this kind of model, in that, the majority of outcomes are typically similar, but the occasion one will have excessive casualties (or conversely, light casualties). This is illustrated in Figure 5 for several values of *k*, for case I.

**Temporal correlation**

If the temporal structure function is as in equations 4 and 5, then the spectrum of the casualty function should exhibit a power law of slope $\beta = -(2F+1)$. However, it is not clear how one analyzes the function $B(t)$ here, since once again the number of Blue entities is small and does not constitute much of a signal to characterize. A simpler approach is to record the time of every casualty in all of the 600 runs, and build a graph of number of casualties from all runs at a given time step. Such a graph appears as in Figure 6, using the recorded time of 250 casualties.

The degree of temporal correlation can be characterized by taking the power spectrum of the data. Figure 7 shows the power spectrum of this casualty time series. Here, to improve the confidence of the spectrum, multiple spectra were made using 250 casualty time records at a time, from a total pool of 600 runs (about 9000 casualty locations in total). These spectra were then averaged. The spectra used the first 256 data points in each time series, to produce a 64-point spectrum. It is notable that the power-law exists on the left hand side, while the right hand side displays a flat, white noise spectrum. This is extremely reminiscent of turbulence power laws, which typically terminate in a flat “dissipation scale” region of the spectrum. It is consistent with an earlier suggestion (Lauren and Stephen, 2001) that disorder is highest on the smallest scales of combat, but that the intermediate scales are neither completely ordered nor disordered (hence the power law, implying no characteristic scale exists, but correlations are evident).

The data can also be checked for multifractal characteristics. For example, if the statistical moments of the data scale, we may expect the relation:

$$\left\langle \left|\Delta B\right|^2 \right\rangle \propto \left(\frac{\Delta t}{T}\right)^{-K(2)} \quad (7)$$

to hold, with $K(2)$ a non-integer. In this case, $\Delta t$ is the resolution at which one examines the time series data, $T$ is the coarsest resolution at which one examines the data, and $\Delta B$ is the number of casualties inflicted in the coarse-grained time step $\Delta t$. Here, we stick to the second-order moment.

Figure 8 shows the scale dependence of this moment, for which the power-law slope is –0.11.

These two power law parameters suggest that the casualty data is fractal and both correlated and intermittent. Given these parameter values, one might for example use methods such as fractal cascades to generate artificial casualty time series. Lauren 2000b discussed the implications of such properties to models with layers of unit aggregation. Generally speaking, such intermittent data will always display scaling of the statistical moments and fat-tailed probability distributions for at least some scales, which must be taken into account in aggregated models. However, the author knows of no existing models that actually do this.

**CONCLUSIONS**

An equation based on fractal dimensions is suggested here for describing attrition on dispersed battlefields. This approach may have more general applications in describing interaction in other complex systems.

The approach assumes an inhomogeneous battlefield, with spatial and temporal correlations between entities. It is well known that multiple interactions by groups of autonomous entities produce power laws, and that appears to be the case for the data presented here.

The fractal equation suggested implies that for a certain ensemble of runs the single-shot kill probability does not affect the rate of attrition (measured in terms of time to reach some casualty level) as strongly as one would expect from the Lanchester equation. This is because the personality rules allow the Red force to cluster, leading to better concentration of firepower, making up for lack of individual firepower.

By definition, runs outside of this ensemble never reach the predetermined level of casualties. Clearly, this below-threshold ensemble is not described by the Lanchester equation.

The results appear to support equation 5 reasonably well. The results are quite similar (though more comprehensive) to earlier results we obtained using another model, called ISAAC (Ilachinski, 2000). ISAAC shares many of the same parameters with MANA, though the actual algorithm for calculating movement is quite different. (Lauren, 1999, 2000a, 2001; Lauren and Stephen, 2001).

The evidence here is interesting because of the power law dependence on the $k$ parameter, rather than for just the temporal and spatial scale.

At the very least, the equations presented here provide a convenient theoretical framework for understanding how dispersed groups of autonomous entities interact. Most probably, a library of functions such as those plotted in Figure 3 and Figure 4 will be necessary in order to understand the behavior of a wide range of dispersed systems.

Another important point is that data that display fractal characteristics often also display fat-tailed probability distributions. That appears to be the case for the results presented here. Importantly, the possession of power laws and fat-tailed distributions were identified as key requirements for combat to be able to be modeled as a statistically scaling system, as suggested in Lauren 2000b.

**ACKNOWLEDGEMENTS**

The author wishes to thank Mr Roger Stephen and Mr Mark Anderson for their efforts in constructing the MANA model and for collating the data for this paper. I would also like to thank Dr Alfred Brandstein for his continuing encouragement and support, which has played a vital role in this research, and the New Zealand Army for its sponsorship.

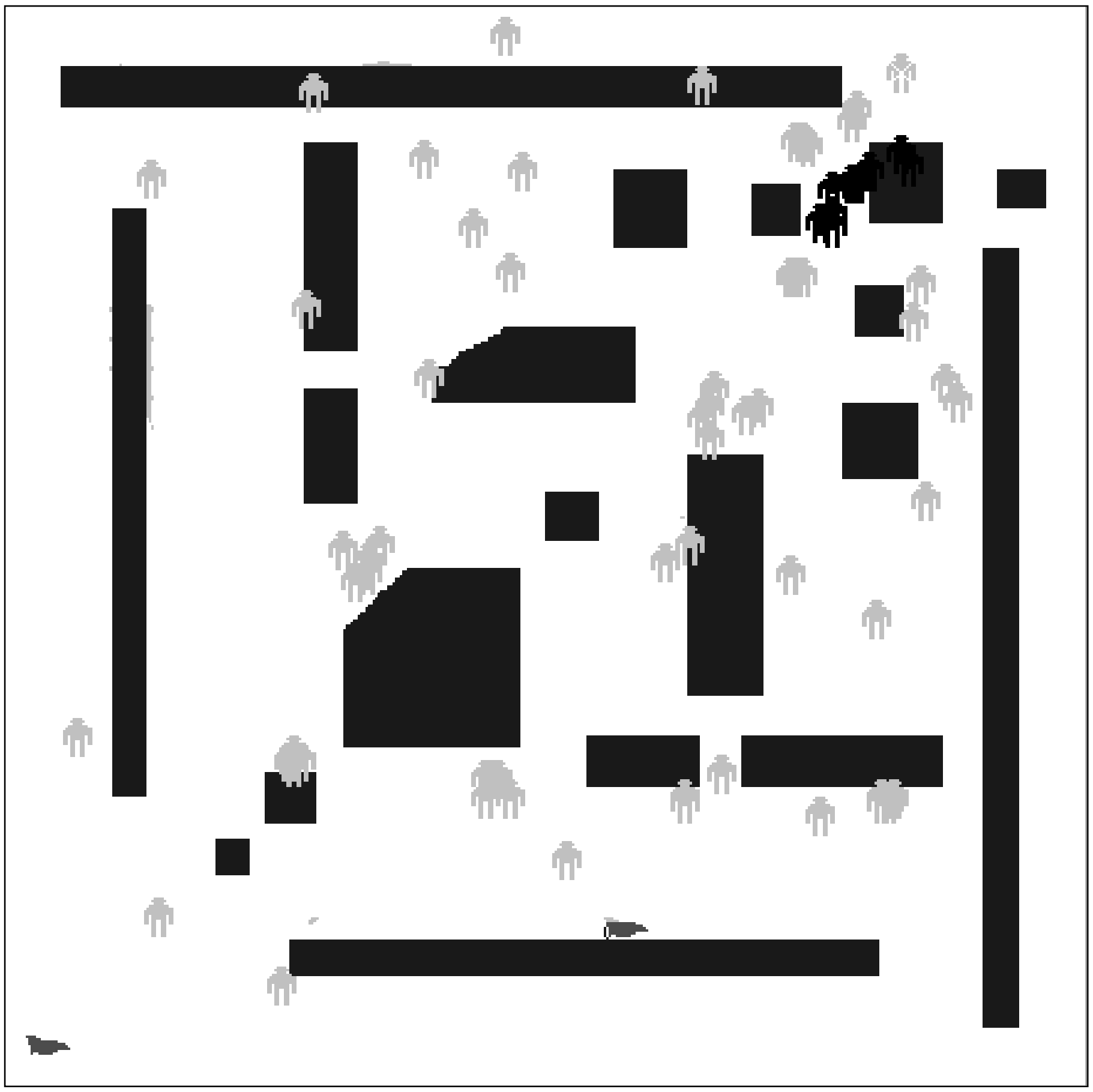

Figure 1: Screen shot of the MANA model scenario analyzed.

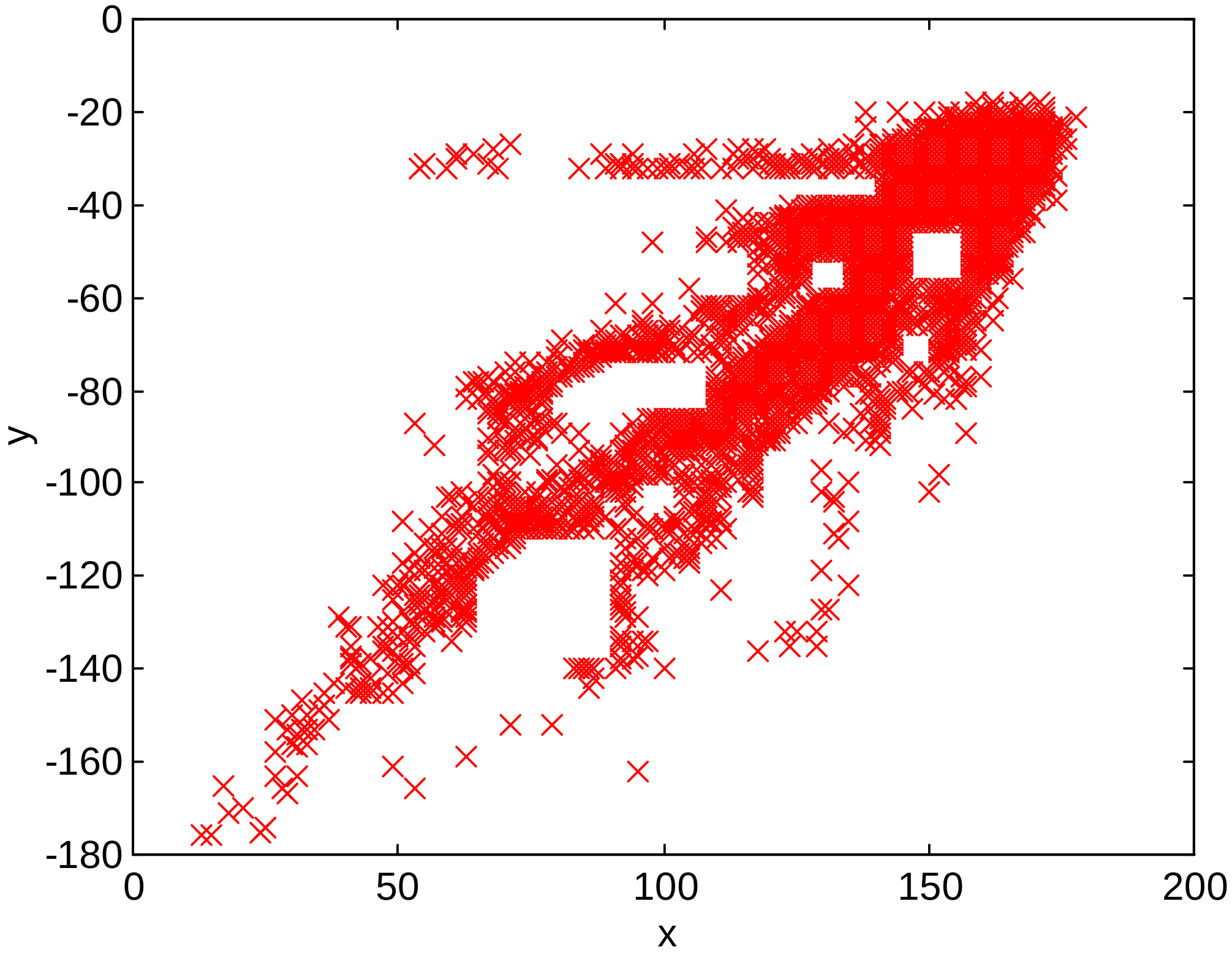

Figure 2: Cumulative distribution of casualty positions from 600 runs for case III.

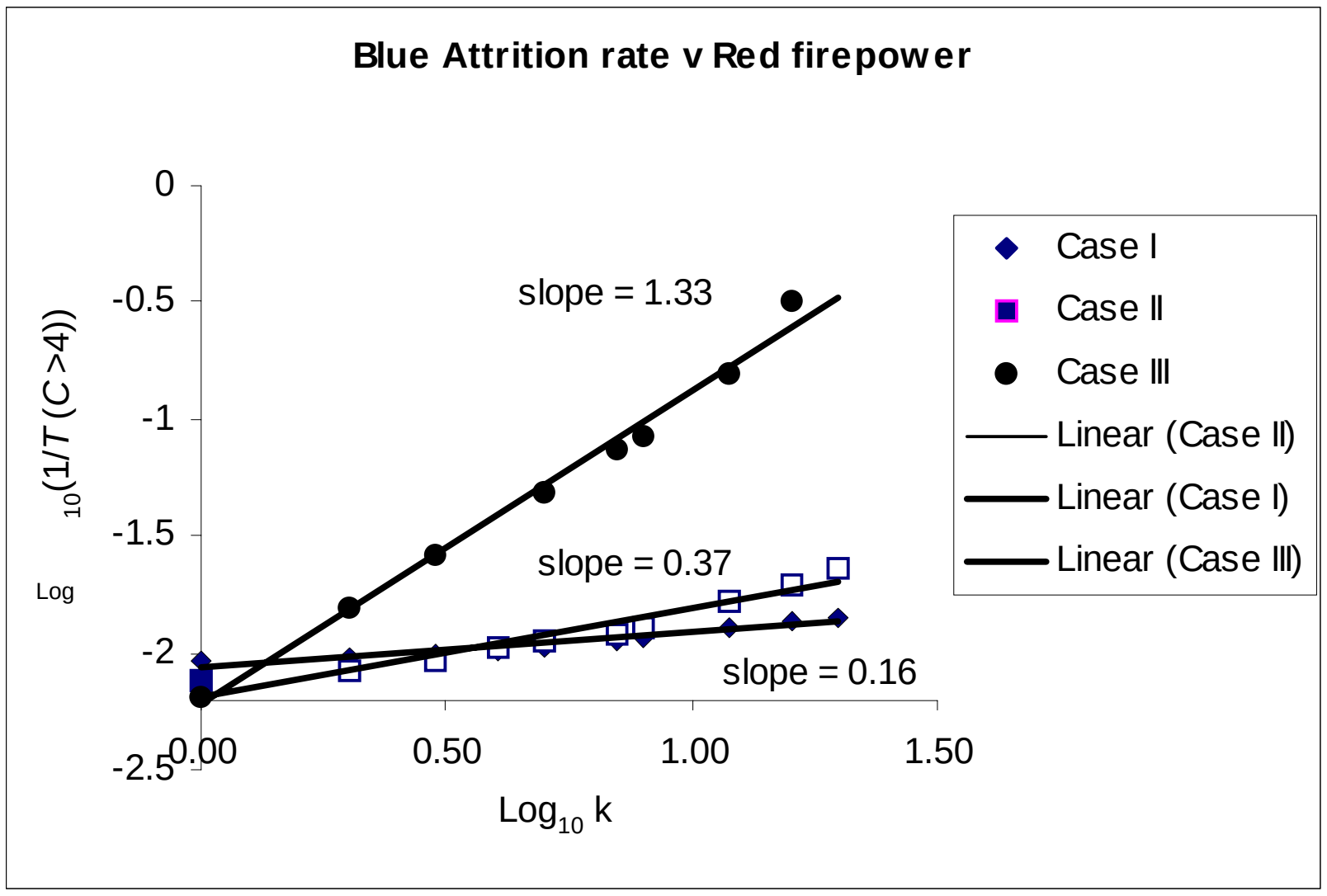

Figure 3: Attrition rate (in terms of 1/time to reach four casualties) as a function of *k*.

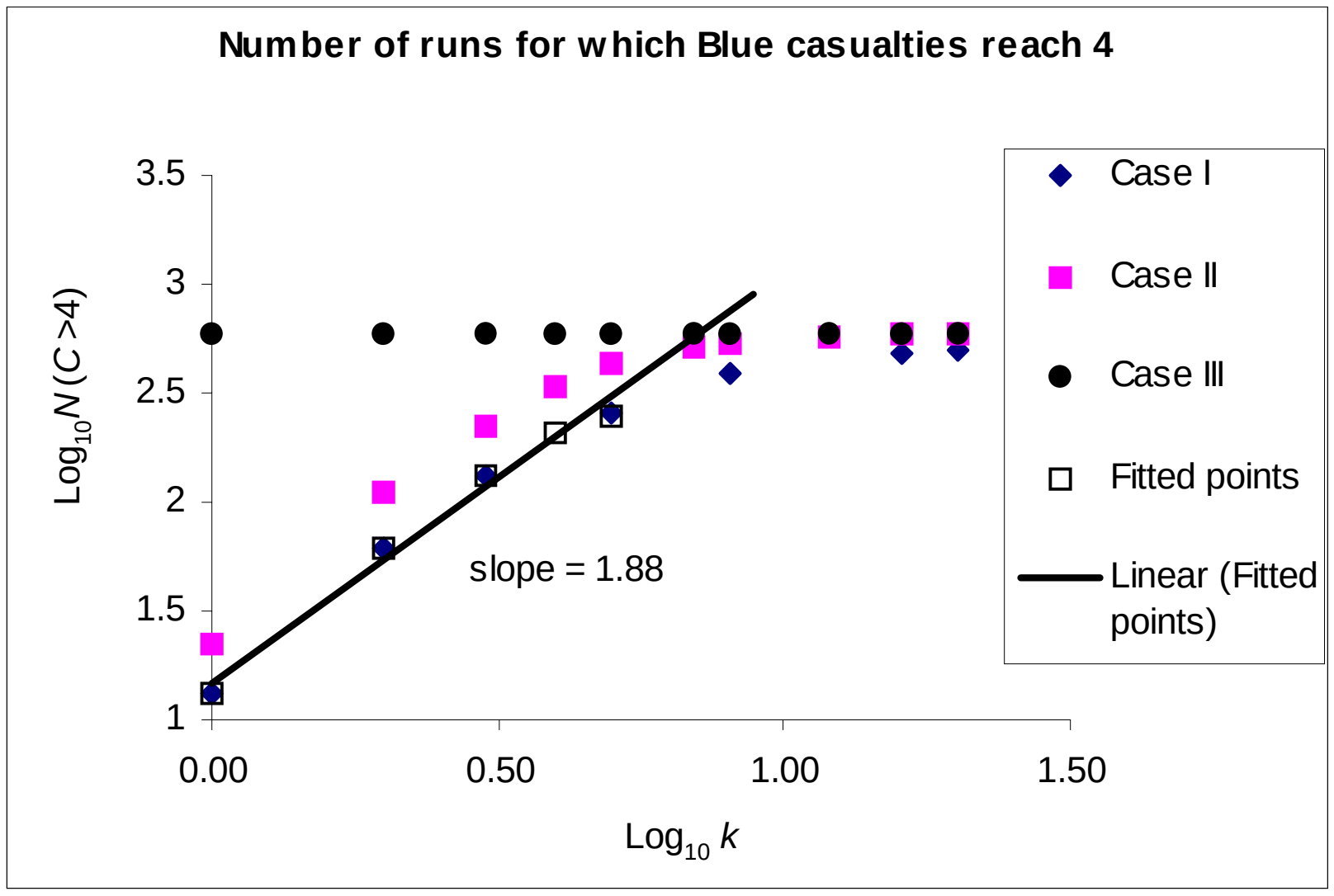

Figure 4: Number of runs (out of 600) that reach four casualties as a function of *k*.

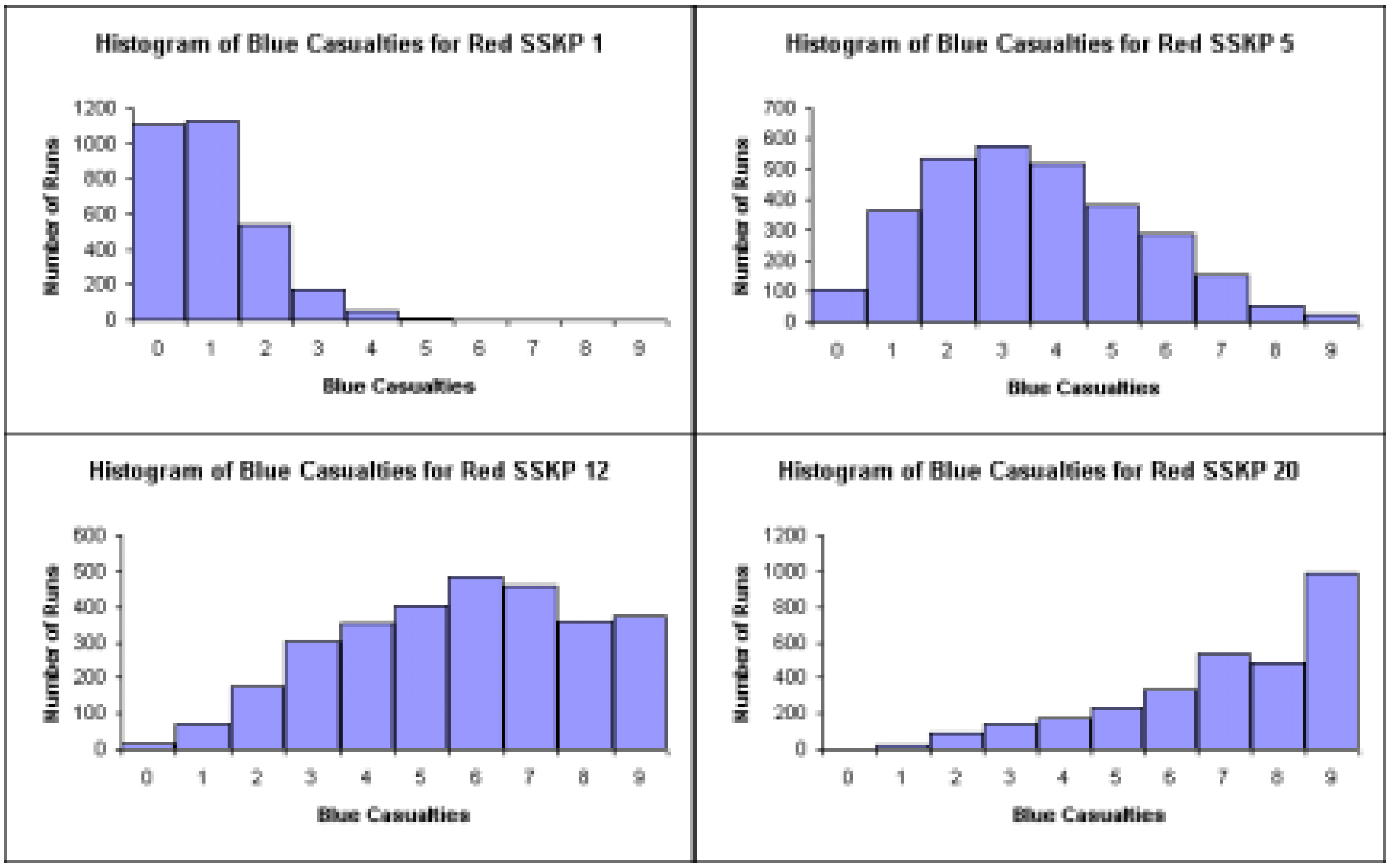

Figure 5: Distribution of casualty levels for case I, as a function of *k*.

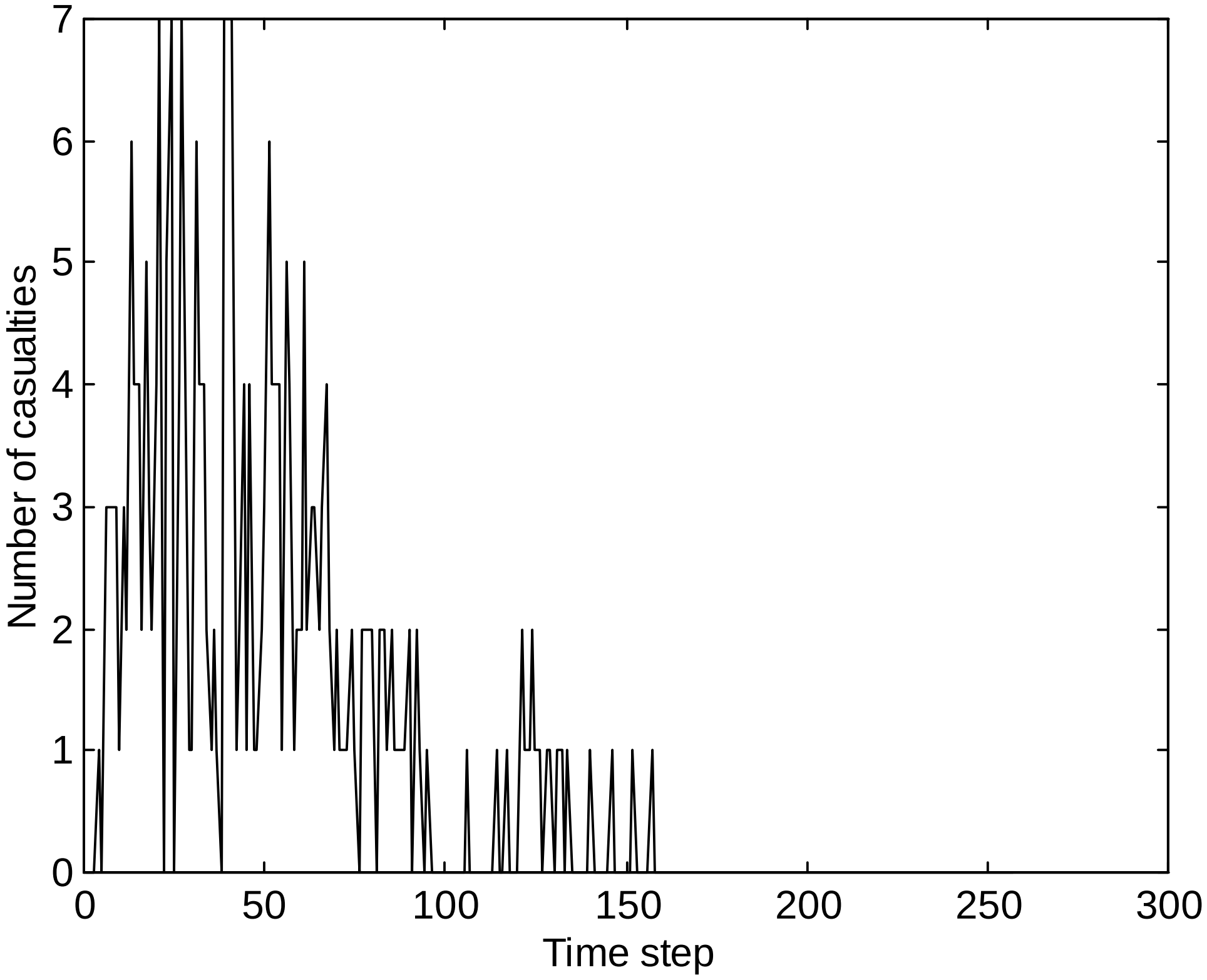

Figure 6: Temporal distribution of 250 casualties obtained from several runs.

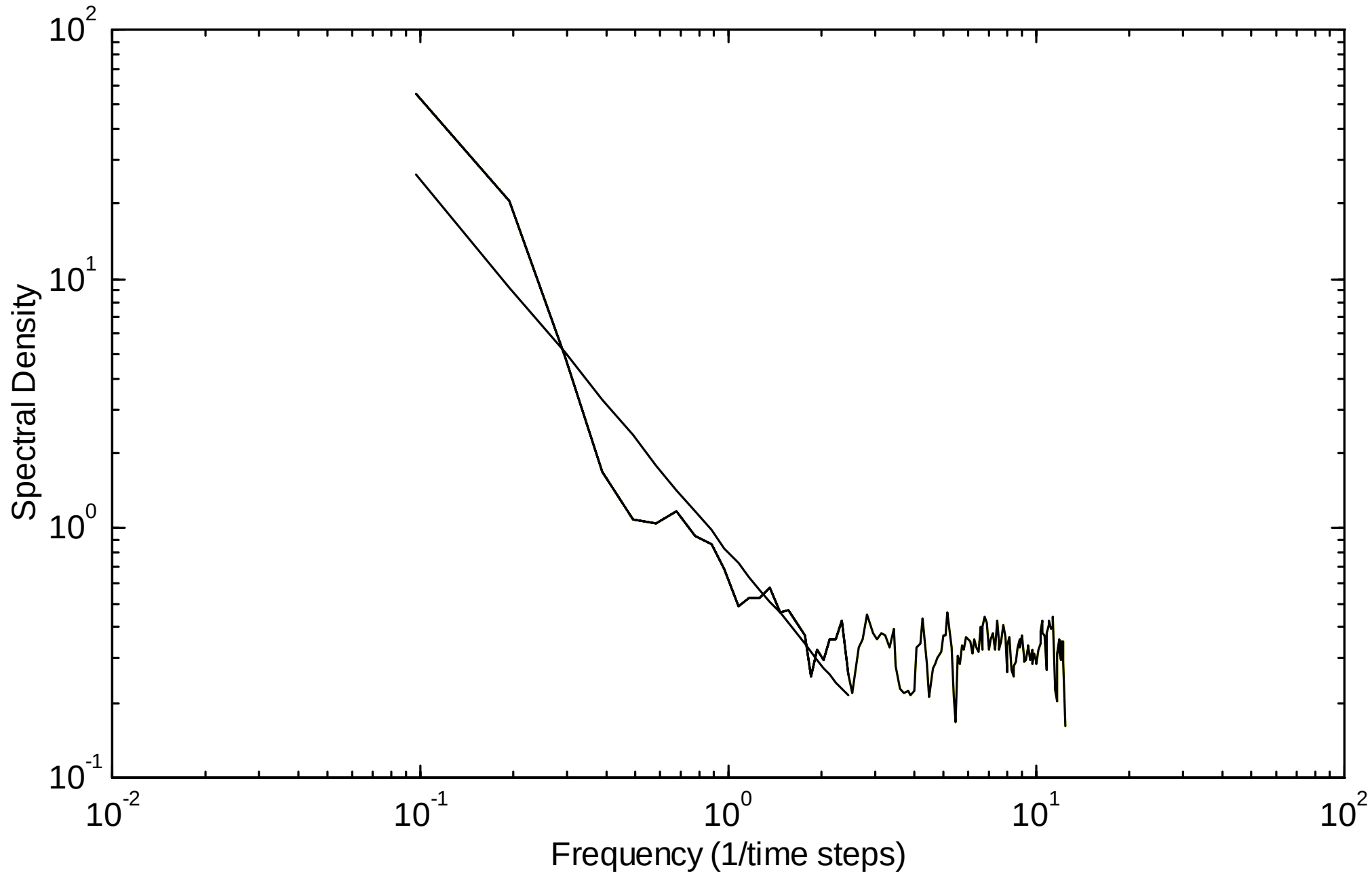

Figure 7: Power spectrum of temporal distribution of casualties. Left-hand side obeys a power law, the right-hand side behaves in a similar way to the turbulent dissipation “white noise” scale.

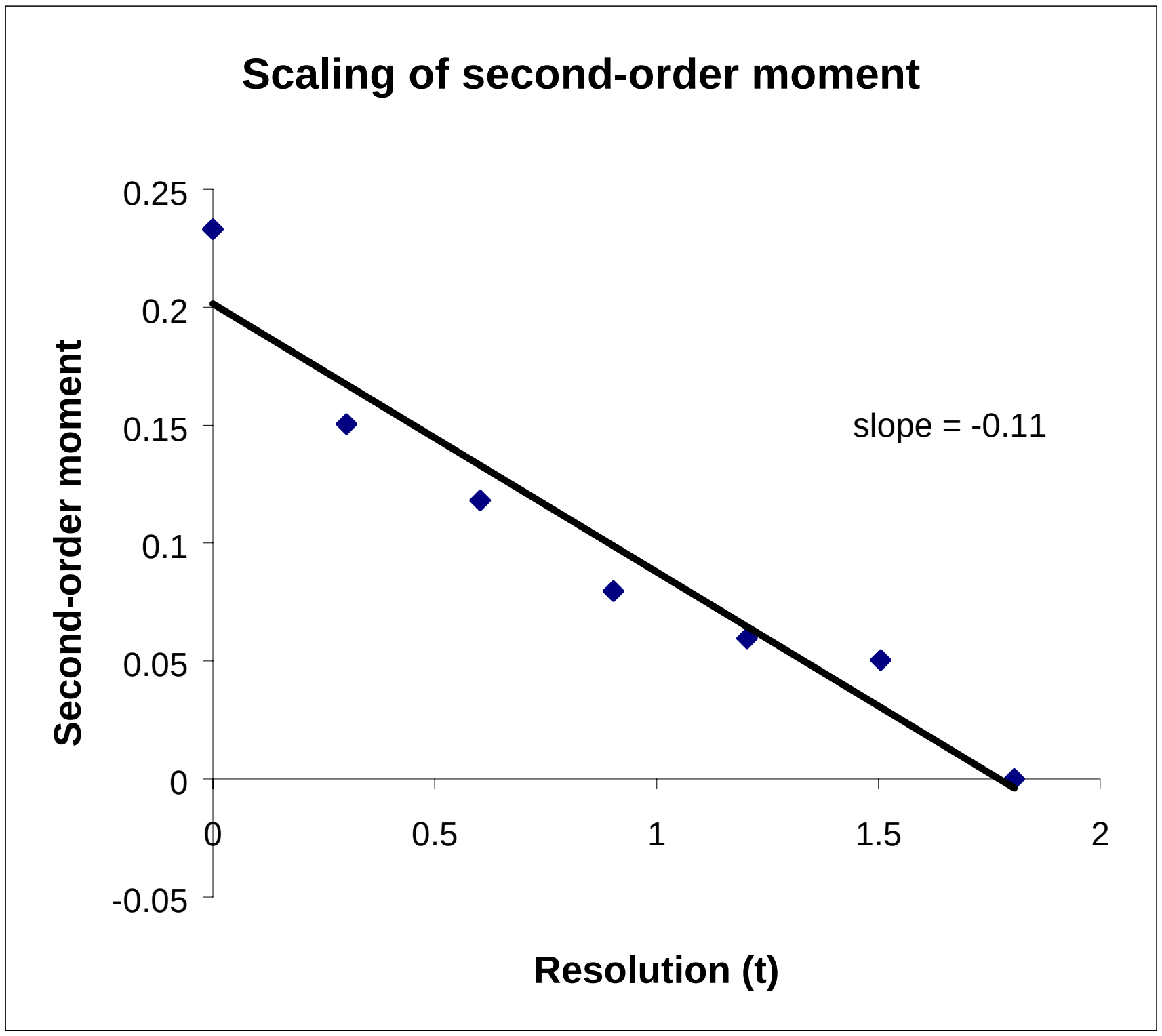

Figure 8: The second-order moment of the casualty time series, as a function of resolution at which the time series is examined.